\documentclass[a4paper,aps,pra,twocolumn,10pt,superscriptaddress,longbibliography]{revtex4-1}
\usepackage{stmaryrd}
\usepackage{}
\usepackage{amssymb}
\usepackage{mathrsfs}
\usepackage{color}
\usepackage{graphicx}
\usepackage{dcolumn}
\usepackage{bm}
\usepackage{hyperref}
\usepackage{bbm}
\usepackage{amsmath}
\usepackage{verbatim}

\newcommand{\appref}[1]{Appendix~\ref{#1}}
\begin{document}


\title{Two-axis-twisting spin squeezing by multi-pass quantum erasure}
\author{Mingfeng Wang}
\affiliation{Department of Physics, State Key Laboratory of Surface Physics and Key Laboratory of Micro and Nano Photonic Structures (Ministry of Education), Fudan University, Shanghai 200433, China}
\affiliation{Department of Physics, Wenzhou University, Zhejiang 325035, China}%
\author{Weizhi Qu}
\affiliation{Department of Physics, State Key Laboratory of Surface Physics and Key Laboratory of Micro and Nano Photonic Structures (Ministry of Education), Fudan University, Shanghai 200433, China}%
\author{Pengxiong Li}
\affiliation{Department of Physics, State Key Laboratory of Surface Physics and Key Laboratory of Micro and Nano Photonic Structures (Ministry of Education), Fudan University, Shanghai 200433, China}%
\author{Han Bao}
\affiliation{Department of Physics, State Key Laboratory of Surface Physics and Key Laboratory of Micro and Nano Photonic Structures (Ministry of Education), Fudan University, Shanghai 200433, China}%
\author{Vladan Vuleti\'{c}}
\affiliation{Department of Physics, MIT-Harvard Center for Ultracold Atoms, and Research Laboratory of Electronics,
Massachusetts Institute of Technology, Cambridge, Massachusetts 02139, USA}%
\author{Yanhong Xiao}
\email{yxiao@fudan.edu.cn}
\affiliation{Department of Physics, State Key Laboratory of Surface Physics and Key Laboratory of Micro and Nano Photonic Structures (Ministry of Education), Fudan University, Shanghai 200433, China}

\date{\today}
\date{\today}

\begin{abstract}
  Many-body entangled states are key elements in quantum information science and quantum metrology. One important problem in establishing a high degree of many-body entanglement using optical techniques is the leakage of the system information via the light that creates such entanglement. We propose an all-optical interference-based approach to erase this information. Unwanted atom-light entanglement can be removed by destructive interference of three or more successive atom-light interactions, with only the desired effective atom-atom interaction left. This quantum erasure protocol allows implementation of Heisenberg-limited spin squeezing using coherent light and a cold or warm atomic ensemble. Calculations show that significant improvement in the squeezing exceeding 10 dB is obtained compared to previous methods, and substantial spin squeezing is attainable even under moderate experimental conditions. Our method enables the efficient creation of many-body entangled states with simple setups, and thus is promising for advancing technologies in quantum metrology and quantum information processing.
\begin{description}
\item[PACS numbers]
42.50.Lc, 03.67.Bg, 42.50.Dv
\end{description}
\end{abstract}

\maketitle
\section{introduction}
 Many-body entangled states of atoms are at the heart of quantum information processing \cite{PhysRevLett.85.5643,RevModPhys.82.1041, NATRUE3} and quantum metrology \cite{PhysRevA.46.R6797,PhysRevLett.82.4619,Nautre1}, of which squeezed spin states (SSS) form an important category. SSS give less quantum fluctuations along a certain direction than the atomic shot noise limit \cite{PhysRevA.47.5138}, and therefore have attracted considerable interest recently since many precision measurements can now reach the atomic shot noise limit. Also, spin squeezing can serve as a criterion to quantify many-body entanglement \cite{PR1}. A widely used approach to create such states is to let atoms interact with a common mode of light, as demonstrated in a variety of systems including cavities \cite{PhysRevA.85.013803,PhysRevA.81.021804,PhysRevLett.104.073602}, cold atoms \cite{PhysRevLett.111.103601,PhysRevLett.102.033601,np1,NATRUE3o} and vapor cells \cite{NATRUE3}. In a majority of current experiments, SSS are conditioned on a measurement. However, from a fundamental point of view, fully determined squeezed states, created in an unconditional way and with maximal entanglement, are still highly desirable. For this, new techniques are needed that allow the complete control of the quantum noise of atoms and light.

A central problem hindering the achievement of a high degree of entanglement by light-mediated atom-atom interaction is the leakage of atomic spin information via the exiting light to the environment, which makes the squeezing process not unitary, and results in mixed states of the atoms with less squeezing. For instance, in the ground breaking proposal on a general spin squeezing scheme pioneered by the Takahashi group \cite{PhysRevLett.94.023003}, the performance is limited by the unwanted atom-light entanglement created during squeezing. One way to erase such entanglement is by using quantum control \cite{PhysRevLett.105.193602}, where a homodyne detection of the probe optical pulse entangled with the atoms followed by feedback control can avoid the information leakage, and thus enhance the amount of achievable squeezing. An alternative approach is based on an optical cavity \cite{PhysRevA.85.013803}. Although these existing erasure protocols can in principle realize a unitary process, they require either near-unity-quantum-efficiency detection or low loss \cite{PhysRevLett.105.193602,PhysRevA.85.013803}, which makes their experimental implementation formidable.

Another outstanding problem in spin squeezing is the lack of simple approaches for unitary two-axis-twisting (TAT) spin squeezing. So far, one-axis-twisting and quantum non-demolition detection (QND) have been realized, but without reaching the Heisenberg limit. As pointed out in Ref. \cite{PhysRevA.47.5138}, TAT promises squeezing to the Heisenberg limit,  but it is challenging to design an experimentally realistic interaction process to realize such a Hamiltonian. Indeed, so far, only a few theoretical proposals exist \cite{PhysRevLett.87.170402,PhysRevA.91.043642,PhysRevA.68.043622,PhysRevLett.107.013601,PhysRevA.65.041803}, and most of them either require special experimental systems such as a Bose-Einstein condensates, Rydberg atoms, or need accurate multiple pulse sequences. Therefore, to date, two-axis-twisted SSS based on entanglement between atoms (instead of internal spin \cite{PhysRevLett.101.073601}) have not been experimentally achieved.

Here, we propose an all-optical interference-based approach to manipulate the quantum noise of atoms and light, which provides a novel yet experimentally feasible way to implement quantum erasure, and enables a simple scheme for TAT spin squeezing. Our method employs the interference of three atom-light interactions, canceling entanglement between the atoms and the output light, but keeping the effective nonlinear interaction between atoms. Such a quantum erasure does not use detection or feedback, is loss tolerant and experimentally feasible. Our scheme does not have special experimental requirements, and only involves an ordinary polarized coherent laser beam under off-resonant Faraday interactions with an ensemble of warm or cold atoms. We note that  other multi-pass schemes have been discussed before for quantum memory applications \cite{PhysRevA.78.010307,RevModPhys.82.1041}, but not for spin squeezing.

\section{ Model and basic interaction}
We consider off-resonant interaction between the collective spin of an ensemble of identical atoms $\textbf{J}=(J_x,J_y,J_z)$ and a coherent light pulse \cite{PhysRevLett.85.5643,RevModPhys.82.1041}. The spin components satisfy the usual angular momentum commutation relations $[J_y,J_z]=iJ_x$, and obey Heisenberg's uncertainty relation $(\Delta J_y)^2\cdot(\Delta J_z)^2\geq|\langle J_x\rangle|^2/4$. For simplicity we assume the ground state atom to be a spin-1/2 system (see Fig. 1). Before interacting, the atomic spins are polarized along the $x$ axis, forming a coherent spin state (CSS) with mean values $\langle J_x\rangle= N_{at}/2, \langle J_y\rangle=\langle J_y\rangle=0$ and variances $(\Delta J_y)^2=(\Delta J_z)^2=|\langle J_x\rangle|/2= N_{at}/4$, as shown in Fig. 2(a)(i).
Such a state will be squeezed according to the Wineland criterion \cite{PhysRevA.50.67} if
\begin{eqnarray}
{\xi ^2} = \frac{{N_{at}{{\left( {\Delta {J_ \theta }} \right)}^2}}}{{{{\left| {\left\langle {{J_x}} \right\rangle } \right|}^2}}} < 1\label{Eq1},
\end{eqnarray}
where ${J_\theta } =  {J_z}\cos \theta -  {J_y}\sin \theta$ with $\theta\in[0,2\pi]$. The input light pulse is
composed of a strong $x$-polarized component with the carrier frequency
$\omega_0=2\pi c/\lambda$ and a weak quantum component polarized
along the $y$ direction. The $y$-polarized component is relevant here, and in the narrow frequency band limit can be described by spatially localized modes \cite{PhysRevA.72.052313} ${x_L}(r) = \frac{1}{{\sqrt {4\pi } }}\int_b {d\omega ({a_y}{e^{ - i({\omega _0} - \omega )r/c}} + H.c.)} ,{p_L}(r) =  - \frac{i}{{\sqrt {4\pi } }}\int_b {d\omega ({a_y}{e^{ - i({\omega _0} - \omega )r/c}} - H.c.)}$ with commutation relations $[x_L(r),p_L(r)]=ic\delta(r-r')$, where the spatial argument $r$ denotes the distance along the optical path,
 $a_y$ is the annihilation operators for $y$-polarized photons, $b$ is the bandwidth of the pulse, $c$ is the speed of light, and the delta function has a width $c/b$. Note that ${x_L}$ (${p_L}$) is in fact proportional to the Stokes vector $S_y$($S_z$) on a Poincar\'{e} sphere \cite{PhysRevLett.85.5643}. The quantum component is initially in a vacuum state such that $\langle x_L(r)\rangle=\langle p_L(r)\rangle=0$, $\langle x_L(r) x_L(r')\rangle=\langle p_L(r) p_L(r')\rangle=ic\delta(r-r')$.
 If the light pulse propagates along the $z$ axis and is tuned far off the atomic resonance, the forward scattering process in a one-dimensional model is described by the Faraday-type Hamiltonian \cite{PhysRevA.72.052313,RevModPhys.82.1041}:
\begin{eqnarray}
V_1=\frac{\hbar\chi}{\sqrt{T}}J_zp_L(r=0)\propto J_zS_z\label{Eq2},
\end{eqnarray}
where the dimensionless coupling constant is given by $\chi^2=\eta\alpha_0/2N_{at}$, with the decay parameter $\eta=N_{ph}\sigma\Gamma^2/A\Delta^2$ and the optical depth on resonance $\alpha_0=N_{at}\sigma/A$. Here $N_{ph}$ is the overall photon number of the pulse with duration T, $\sigma$ is the scattering cross section, $\Gamma$ is the natural linewidth (HWHM) of the atomic transition, $A$ is the effective beam cross section, $\Delta$ is the detuning
from the optical transition, and the sample is assumed to be located at $r=0$.

The key physics of the Hamiltonian $J_zS_z$ (\ref{Eq2}) can be summarized as follows: For light propagating along $z$, the Stokes vector precesses around the $z$-axis with a rate proportional to the atom spin $J_z$. If the light is $x$-polarized ($S=S_x$), then its $S_y$ component will pick up the atoms' $J_z$ information. Similarly, if atoms are polarized along $x$, under evolution governed by this Hamiltonian, its $J_y$ component will pick up the light's $S_z$ information. Therefore, the atom and light act as a quantum data bus for each other.

\section{ Working principle of quantum erasure}

\begin{figure}[b]
\centering
\includegraphics[scale=0.4]{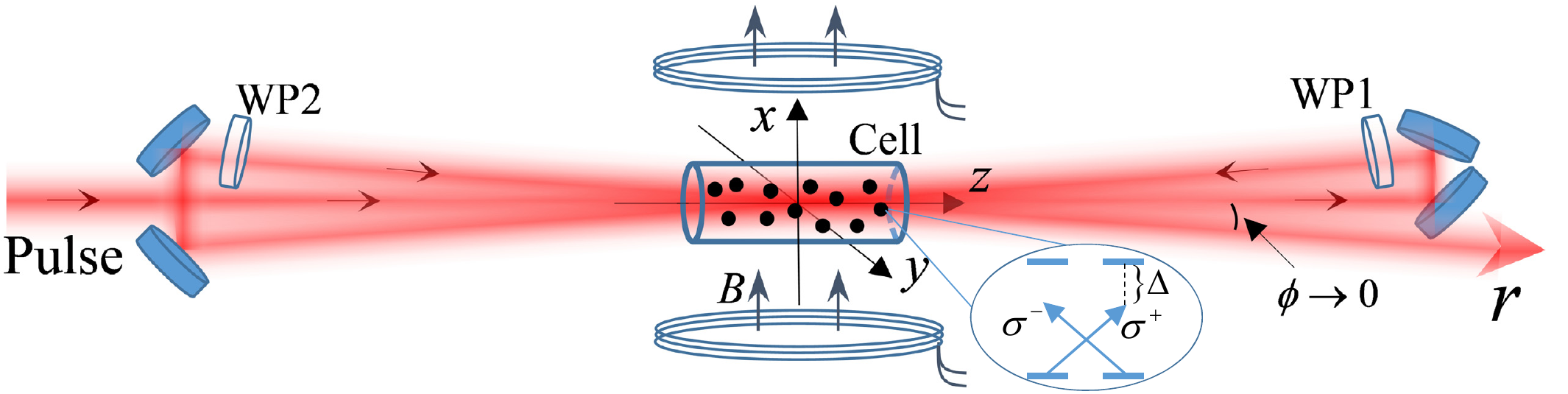}
\caption{ Schematic of the proposed setup for two-axis-twisting spin squeezing. A light pulse passes three times through an atomic
ensemble placed in a magnetic field, with waveplates in between the passes. The light exiting the system contains no information about the atomic spin, and the atomic spin is perfectly squeezed. See text for details.}
\end{figure}

With the above physical picture, the principle of our three-pass interference-based quantum erasure can be understood. As shown in Fig.1, an $x$-polarized light pulse propagates along $z$ and passes through the atomic sample three times, with its quantum components (in the $y$-$z$ plane of the Poincar\'{e} sphere) rotated by two waveplates (WP1,2) between passages. From the Hamiltonian (\ref{Eq2}), one readily derives, for a single pass, the input-output relations \cite{RevModPhys.82.1041}:
$x_L^{out}  = x_L^{in}  + \chi J_z^{in}, p_L^{out}=p_L^{in},J_y^{out}  = J_y^{in}  + \chi J_xp_L^{in},J_z^{out}=J_z^{in}$ , where $x_L^{in, out}$ and $p_L^{in, out}$ are the normalized light quadrature defined as $
q_L^{in}  = (1/\sqrt T )\int_0^T {d\tau q_L (-c\tau,0)} ,q_L^{out}  = (1/\sqrt T )\int_0^T {d\tau q_L (cT-c\tau,T)}$ with $q\in\{x,p\}$, and for the atomic spin $\textbf{J}^{in}=\textbf{J}(0),\textbf{J}^{out}=\textbf{J}(T)$. One can see that after one interaction, the $x$-quadrature ($S_y$) of the outgoing field now carries information about $J_z$. Then, we assume that the Stokes vector is rotated by WP1 around the $x$-axis before interacting with the sample again. If the polarization rotation is $90$ degrees, then after the second pass, $J_y^{out'}  \simeq J_y^{in}  + \chi J_x (x_L^{in}  + p_L^{in} ) + \chi ^2 J_xJ_z ,J_z^{out'}  = J_z^{in}$. At this stage, the result of the two-passage interaction is exactly the same as the previous double-pass (DP) scheme that realizes non-unitary one-axis-twisting squeezing\cite{PhysRevLett.94.023003}. On the one hand, light picks up the atomic information $J_z$ and imprints them onto $J_y$, inducing a one-axis-twisting (OAT) nonlinear spin dynamics represented by the $J_xJ_z$ term; but on the other hand, light leaves its quantum noise (that is, $x_L+p_L$) imprinted on the atoms, producing unwanted entanglement between light and atoms, which prevents the realization of an ideal one-axis twisted state and greatly reduces the amount of achievable squeezing \cite{PhysRevLett.94.023003}. An immediate solution is to perform a projection measurement of the exiting light followed by an electronic feedback control \cite{PhysRevLett.105.193602}. Instead, here we propose to disentangle the collective spin and light in a purely optical way by adding a third interaction. The essential idea is to appropriately adjust the light polarizations (only their quantum components) between passes so that the spin-light entanglement created by all passes destructively \emph{interferes}, leaving only the net effect---a nonlinear OAT of the collective spin, $\propto J_z^2$. Therefore, the quantum erasure is built into the interaction.

We describe the choice of the polarization rotation angles between passes using the vector plots [Fig.2(b)], where we consider the evolution of a point in phase space (also on the Poincar\'{e} sphere) for light.  From the light's point of view, it experiences the atom' collective spin successively for three times, and according to the $J_zS_z$ Hamiltonian, atomic information equal to $\chi J_z$ (solid red arrow) is added to $x_L$ ($\propto S_y$) each time, as depicted separately by the three plots of Fig.2(b). The second pass picks up noise $-\chi J_z$ since light propagation direction is reversed. If the Stokes vector is rotated by 60 degrees between each interaction, then after the third interaction, the three solid red arrows add up to zero [Fig.2(b)(iii)], which means the light exits without containing any information about the atomic spin. From the atoms' point of view, in each pass, its $J_y$ component obtains noise from the $S_z$ component ($p_L$) of light, as indicated by dashed arrows on the $p_L$ axis in Fig.2(b). If we sum up all the $p_L$ ($S_z$) vectors from the three figures in Fig.2(b), with a minus sign for the second passage (again due to reversed light prorogation direction), it can be seen that only the component proportional to $\chi J_z$ [the red dashed arrow in (ii)] is left. This fact that the atoms' $J_y$ component is displaced by an amount of $\frac{\sqrt{3}}{2}\chi J_z$ indicates exactly the dynamics governed by the $J_z^2$ OAT Hamiltonian.

 \begin{figure}[t]
\centering
\includegraphics[scale=0.305]{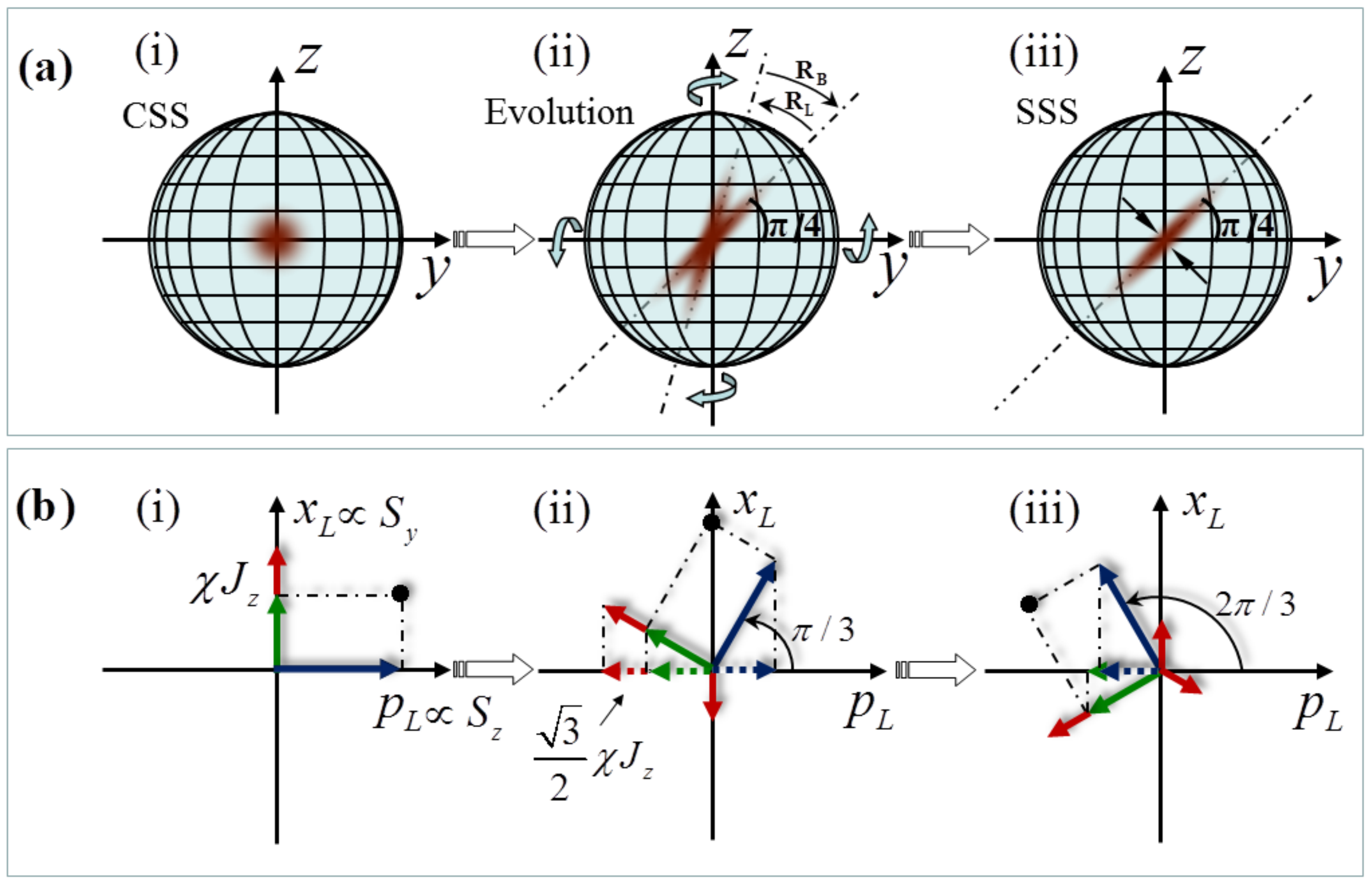}
\caption{(color online). (a) Illustration of the spin distributions evolution on the Bloch sphere for the proposed triple pass scheme. The atomic state is initially prepared in (i) the  coherent spin state. (ii) The atom-light interaction induces two effects. First, it squeezes the spin uncertainty along the $3\pi/4$ direction. Second, it rotates the transverse spin around the $x$ axis at a rate $R_L$, which can be compensated by the Larmor precession $R_B$, leading to the net effect (iii). (b) Phase-space portrait of the light's evolution induced by the spin-light interactions. (i) During the $V_1$ interaction, $x_L$ ($\propto S_y$) picks up atomic noise $J_z$. (ii) After a rotation produced by WP1 of the phase space by an angle $\pi/3$, the $V_2$ interaction imprints atomic noise $-J_z$ onto the $x_L$ quadrature. (iii) After another $\pi/3$ rotation by WP2, $x_L$ `sees' $J_z$ again. The sum of the three atomic spin contributions to the polarization of the light is zero, thereby canceling the spin-light entanglement. Dashed arrows are projections of all the vectors onto the $p_L$ axis as explained in the text.}
\end{figure}

\section{ Implementation of a unitary two-axis twisting (TAT) spin squeezing}
The above scheme for the unitary OAT Hamiltonian $J_z^2$ can be extended to a TAT spin squeezing scheme. First, one can write $J_z^2=(J_z^2-J_y^2)/2+(J_z^2+J_y^2)/2$, where the first term refers to the TAT transformation that creates pure squeezing with exponential growth along the $3\pi/4$ direction of the $XY$ plane [Fig. 2(a)(ii)], and the second term makes the transverse spin components rotate around the $x$ axis (denoted by $R_L$), which causes the unwanted swirling effect \cite{PhysRevA.47.5138}. Next, the spin dynamics induced by this second term is ${{\dot J}_y} = ({J_x}{J_z} + {J_z}{J_x})/2 \simeq {N_{at}}{J_z}/2,{{\dot J}_z} =  - ({J_x}{J_y} + {J_y}{J_x})/2 \simeq  - {N_{at}}{J_y}/2$, which indicates $(J_z^2+J_y^2)/2\simeq -N_{at}J_x/2$. Therefore, a constant magnetic field in the $x$-direction can impose an opposite Larmor precession $R_B$ with respect to $R_L$, and thus cancel the light-induced rotation, leading to a pure TAT spin state [Fig. 2(a)(iii)].

We now describe formally the creation of the two-axis-twisted spin state. Our TAT scheme (Fig.1) relies on the near simultaneous passage of a light pulse for three times through the atomic sample, placed in a homogenous magnetic field. The pulse's spatial length is assumed to be much longer than the loop length between mirrors so that the laser pulse encounters itself in the cell. Such overlap is key to the current scheme, since (see below) it allows a continuous light-induced rotation $R_L$, which can then be instantaneously eliminated by the magnetic field, enabling a one-step realization of TAT. We note that pulse overlap is not necessary for the above unitary OAT scheme.
After the first pass, WP1 (with optical axis along $x$-direction) introduces a relative phase shift $\alpha$ between the $x$- and $y$-polarization component, giving rise to a rotation of optical quadratures $x_L$ and $p_L$ [Fig.1(b)(ii)] around the $x$-axis: $x_L\rightarrow x_{L,\alpha}= x_L\cos\alpha+p_L\sin\alpha,p_L\rightarrow p_{L,\alpha}= p_L\cos\alpha-x_L\sin\alpha$. Next, the beam is reflected back into the sample, at a small angle $\phi$ with respect to $-z$ direction. In the limit $\phi\rightarrow 0$,  one approximately obtains the interaction: $V_2=-\frac{\hbar\chi}{\sqrt T} J_zp_{L,\alpha}(d_1)$, where the spatial argument reflects the fact that this interaction happens after the pulse has traveled some distance $d_1$ in the loop between the mirrors and the minus sign stems from the change of the light propagation direction. Then, after passing the WP2 (also with optical axis along $x$-direction) and traveling a distance $d'$, the third interaction $V_3=\frac{\hbar\chi}{\sqrt T} J_zp_{L,\beta}(d_2)$ happens, where $d_2=d_1+d'$ and $\beta=\alpha+\alpha'$ with $\alpha'$ the relative phase shift induced by WP2. Altogether, the triple-pass interaction can be described by $H=H_A+H_L+V_1+V_2+V_3$, where $H_A=\hbar\Omega J_x$ refers to the precession of $J_y$ and $J_z$ around the $x$-axis with frequency $\Omega$ due to a magnetic field, and $H_L$ denotes the Hamiltonian for the free-space radiation field. From this Hamiltonian, one may evaluate the Heisenberg equations for light and atoms, yielding the following equations of motion (see \appref{Evolution of the spin state} for details):
\begin{eqnarray}
 \frac{d}{{dt}} J_y \left( t \right)&=&-\Omega J_z\left( t \right)+ \frac{\chi }{{\sqrt T }} \langle J_x\rangle\left[ {p_L \left( {0,t} \right)} \right.\nonumber\\
 &&\left. { -p_{L,\alpha } \left( { d_1,t} \right)+ p_{L,\beta } \left( {d_2,t} \right)} \right], \label{Eq3}\\
 \frac{d}{{dt}}J_z \left( t \right) &=& \Omega J_y\left( t \right), \label{Eq4}\\
 \left(\partial _t+c\partial _z\right) x_L \left( {r ,t} \right) &=& \frac{{c\chi }}{{\sqrt T }}J_z \left( t \right)\left[ {\delta \left( {r} \right)- \delta \left( {r - d_1 } \right)\cos \alpha} \right. \nonumber\\
 &&\left. { + \delta \left( {r  - d_2 } \right)\cos \beta } \right], \label{Eq5}\\
 \left(\partial _t+c\partial _z\right) p_L \left( {r,t} \right) &=& -\frac{{c\chi }}{{\sqrt T }}J_z \left( t \right)\left[ {\delta \left( {r - d_1 } \right)\sin \alpha } \right. \nonumber\\
 &&\left. { + \delta \left( {r  - d_2 } \right)\sin \beta } \right]\label{Eq6},
 \end{eqnarray}
 where we have assumed that the spin-light coupling is weak, so that the spin orientation does not deviate much from the $x$-direction during the interaction. Under this assumption one may omit the time evolution of the $x$-component and replace $J_x$ by its average value.  Equation (\ref{Eq3}) has a clear interpretation: At a certain instant in time $t$, the collective spin $J_z$ simultaneously receives information from three different light spatial modes, which reflects the fact that these spatial modes overlap in the cell.
 To solve the atomic equation (\ref{Eq3}), we first solve the Eqs. (\ref{Eq5}) and (\ref{Eq6}) for light, and then substitute the results into Eq. (\ref{Eq3}), yielding
\begin{eqnarray}
 \frac{d}{{dt}}J_y \left( t \right) &=&  - \Omega J_z \left( t \right) + \frac{{\chi ^2 }}{T}\langle J_x \rangle \left[ {\sin \alpha J_z \left( {t - d_1 /c} \right)} \right. \nonumber\\
  &&- \sin \left( {\alpha  - \beta } \right)J_z \left( {t - d_2 /c + d_1 /c} \right) \nonumber\\
 &&\left. { - \sin \beta J_z \left( {t - d_2 /c} \right)} \right] + \frac{\chi }{{\sqrt T }}\langle J_x \rangle {\cal F}_L \left( t \right), \label{Eq7} \end{eqnarray}
with $\mathcal {F}_L= (1 - \cos \alpha  + \cos \beta ) p_L (-ct,0) - (\sin \beta  - \sin \alpha ) x_L (-ct,0)$. The terms in the square brackets denote the atomic information brought back by light, whose time arguments reflect the time when the collective spin $J_z$ is interacting with the respective light field. The last term of Eq. (\ref{Eq7}) indicates that the light leaves its information $\mathcal {F}_L$ in the sample, which, as analyzed above, is unwanted for spin squeezing. However, if WP1 and WP2 are both $\lambda/6$ wave plates, we have $\alpha=\pi/3,\beta=2\pi/3$, which means $\mathcal {F}_L=0$. This is consistent with the intuitive plots in Fig.2. Moreover, to further simplify Eq. (\ref{Eq7}), we make the experimentally feasible assumption that $d_{1,2}/c\ll 1/\Omega$, which means that the elapsed time during the light running in the loop is much shorter than the Larmor period \footnote{This assumption is feasible, since $d_{1,2}$ in reality is normally of the order of ns, while $1/\Omega$ involved here is of the order of ms.}. Under this assumption, we approximately have $J_z(t-d_{1,2}/c)\simeq J_z(t-d_2/c+d_1/c)\simeq J_z(t)$ and finally obtain
\begin{eqnarray}
\frac{d}{{dt}}\left( \begin{array}{l}
 J_y \left( t \right) \\
 J_z \left( t \right) \\
 \end{array} \right) = \left( {\begin{array}{*{20}c}
   0 & {\frac{{\sqrt 3 \kappa ^2 }}{{2T}}  - \Omega }  \\
   \Omega  & 0  \\
\end{array}} \right)\left( \begin{array}{l}
 J_y \left( t \right) \\
 J_z \left( t \right) \\
 \end{array} \right),\label{Eq8}
\end{eqnarray}
where we have defined the dimensionless coupling constant $\kappa^2=\chi^2\langle J_x\rangle=N_{ph}N_{at}(\sigma\Gamma/A\Delta)^2$.

Next,  we let $
\Omega=\sqrt 3 \kappa ^2 /4T $, which means that the angular velocity of the transverse spin induced by the magnetic field
is the same as the angular velocity caused by the OAT dynamics, but with opposite directions. Under this condition, Eq. (\ref{Eq8}) can be directly solved to yield
\begin{eqnarray}
J_{\pi /4}^{out}  = e^{\frac{{\sqrt 3 \kappa ^2 }}{4}} J_{\pi /4}^{in} ,J_{3\pi /4}^{out}  = e^{ - \frac{{\sqrt 3 \kappa ^2 }}{4}} J_{3\pi /4}^{in},\label{Eq9}
\end{eqnarray}
which represents the main result of this paper. It is evident that, a pure TAT transformation $U_{TAT}=e^{-i\sqrt{3}\kappa^2(J_y^2-J_z^2)/8}$ is successfully applied to the spin state. Spin fluctuations are now squeezed along the $\theta=135^{\circ}$ direction of the transverse spin at a rate that shrinks them exponentially, yielding the squeezing parameter [Eq. (\ref{Eq1})] $\xi^2=\exp(-\sqrt{3}\kappa^2/2)$. Compared to the DP scheme whose squeezing parameter is ${\xi^2 _{DP}} = 1 + ({\kappa ^4}/2 + {\kappa ^2})[1 - \sqrt {1 + 4/{{(2 + {\kappa ^2})}^2}} ] \Rightarrow {\lim _{\kappa  \to \infty }}2/{\kappa ^2}$ \cite{PhysRevLett.105.193602}, our scheme strongly enhances the degree of squeezing.
 \begin{figure}[t]
\centering
\includegraphics[scale=0.53]{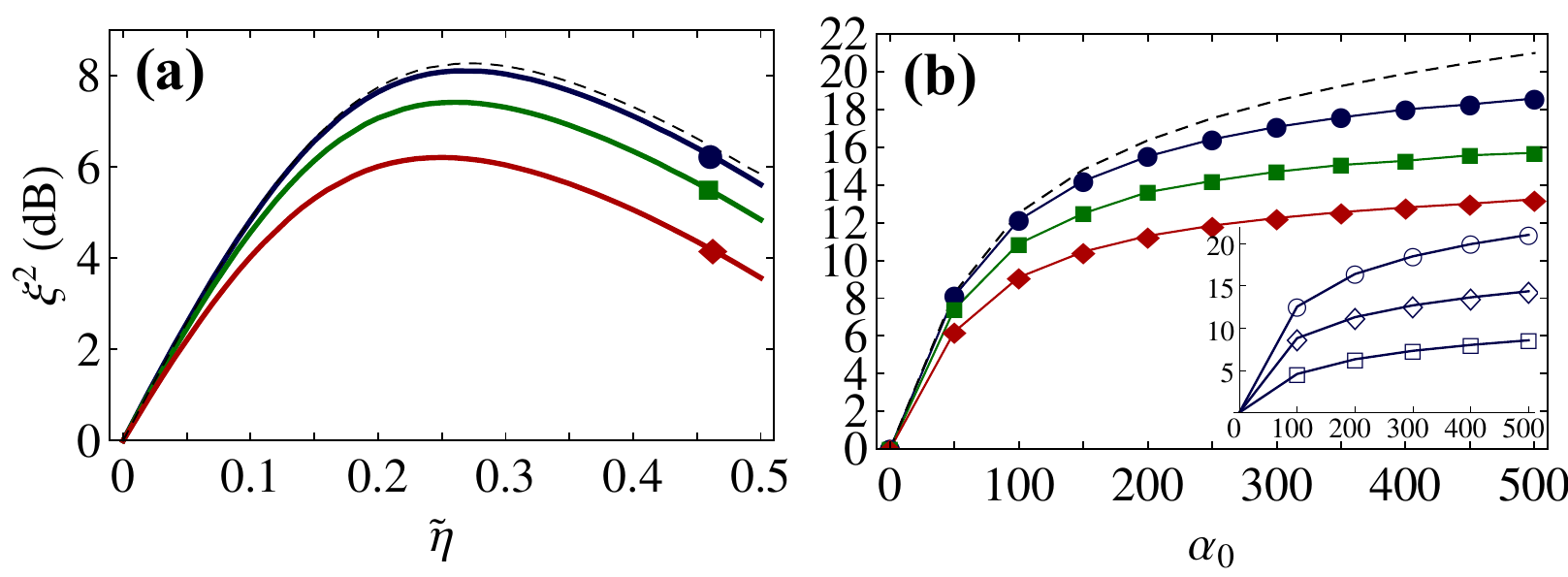}
\caption{Calculated squeezing vs. photon number (expressed in terms of atomic decay $\tilde\eta$) and optical depth $\alpha_0$, for three different values of photon loss.  (a) The optimal squeezing $\xi^2$ vs total photon number for $\phi=0.05$ and $\alpha_0=50$. (b) Peak squeezing vs optical depth $\alpha_0$. Lines with symbols from top to bottom: no optical loss $\zeta=0\%$(blue {\large\textcolor[rgb]{0.00,0.00,0.50}{\textbullet}}), $\zeta=2\%$ (green {\tiny\textcolor[rgb]{0.00,0.50,0.00}{$\blacksquare$}}), $\zeta=6\%$ (brown {\footnotesize\textcolor[rgb]{0.50,0.00,0.00}{$\blacklozenge$}}), $\phi=0.05$. The dashed lines denote the case of no optical loss and $\phi=0$. The inset shows the performance of different ideal spin squeezing protocols versus optical depth, from bottom to top: double pass in \cite{PhysRevLett.94.023003} ({\footnotesize{\textcolor[rgb]{0.00,0.00,0.50}{$\oblong$}}}), our OAT ({\large\large{\textcolor[rgb]{0.00,0.00,0.50}{$\diamond$}}}), our TAT ({\Large\textcolor[rgb]{0.00,0.00,0.50}{$\circ$}}).}
\end{figure}

\section{Imperfections}
So far, we have neglected the imperfections of the scheme. In deriving Eq. (\ref{Eq2}), we have assumed $\phi\rightarrow0$ for the angle between the beams. Although in principle, the angle $\phi$ can be made arbitrarily small by, e.g., prolonging the length $d_{1,2}$ of the loop, it is a finite number in reality. As a result, the $p_L$ quadrature will see not only the spin $J_z$ but also $J_y$ during the second and the third passages, which effectively transforms the interaction into: $
V_2  \to V'_2  \propto J_{\pi  + \phi } p_{L,\alpha } ,V_3  \to V'_3  \propto J_{2\pi  - \phi } p_{L,\beta }
$. Also, atomic decay from the weakly populated excited state causes random rotation of the ground state spin, and shortens the total spin: $\langle J_x\rangle\rightarrow \langle J_x\rangle (1-\tilde\eta)$, where $\tilde\eta\approx3\eta$, with the factor of 3 from the light overlap in the atomic sample and $\eta$ proportional to the intensity of input optical field. We neglect spin decoherence from wall collisions (see next paragraph). Correspondingly, the transverse spin components now evolve as
$
\dot J_i = i[H,J_i]/\hbar  - \tilde \eta J_i/2T + \sqrt {\tilde \eta /T} f_{J_i}$ with $i\in\{y,z\}$, where $f_{J_i}$ represents the Langevin noise operators with zero mean and $\langle f_{J_i}(t),f_{J_i}(t')\rangle=\langle J_x\rangle\delta(t-t')/2$.

Furthermore, the light is subject to loss. The above decay event acts as a source of absorption or decoherence for light and reduces the probe photons by the factor $\epsilon=N_{at}\eta/N_{ph}$ \cite{PhysRevA.70.052324}. Another loss mechanism is reflection off cell walls due to their finite reflectivity $r_0$. Leaving and re-entering the cell through one window gives reflectivity $r=2r_0$. The loss effect can be modeled as a beam-splitter type admixture of vacuum components, which transforms the quadratures at $r=d_n$ ($n=1,2$) into $
q(d_n ) \to \sqrt {1 - n\zeta} q(d_n ) + \sqrt {n\zeta} f_{Lq ,3n}$ \footnote{Reflection losses by the first wall is neglected since it can be compensated by using a more intensive coherent state. Also, for simplicity we set $\epsilon\sim r_0$.}, where $\zeta=\epsilon+2r_0$ denotes the overall loss rate caused by crossing the sample through two cell walls, and $
f_{Lq ,3n}  = \sum\nolimits_{i = 1}^{3n} {f_{Lq}^i } /\sqrt {3n}$
 represents the vacuum noise operator with $
{f_{Lq}^i }
$ the Langevin operator of light admixed during the $i$th crossing.

After considering all above imperfections (see \appref{Imperfection of the scheme} for details) and optimizing the protocol with respect to the magnetic field and the relative phases introduced by WPs, we obtain curves showing (Fig.3)
 the squeezing $\xi^2$ v.s. photon number (drawn as the decay parameter $\tilde\eta$) for different photon loss, with $\phi=0.05$ and an optical depth (OD) $\alpha_0=50$. As shown, for each photon loss $\zeta$, there exists a maximal squeezing, achieved at a certain photon number ($\propto \tilde\eta$). Strong squeezing of more than $7.4$ dB is achievable for $\zeta=2\%$. For $\zeta=6\%$, a very respectable squeezing of $6.2$ dB can still be seen. The peak squeezing as a function of $\alpha_0$ for different light losses is plotted in Fig. 2(b). For an atomic system with large optical depth of $\alpha_0= 100$, a degree of squeezing created should be as high as $10.9$ dB for $\zeta=2\%$. A comparison of the performances of different protocols [inset in Fig. 3(b)] shows that our schemes greatly enhance the amount of achievable squeezing. The increase in squeezing compared to the scheme without quantum erasure \cite{PhysRevLett.94.023003} is about 10 dB for an optical depth of 500.
  \begin{figure}[t]
\centering
\includegraphics[scale=0.315]{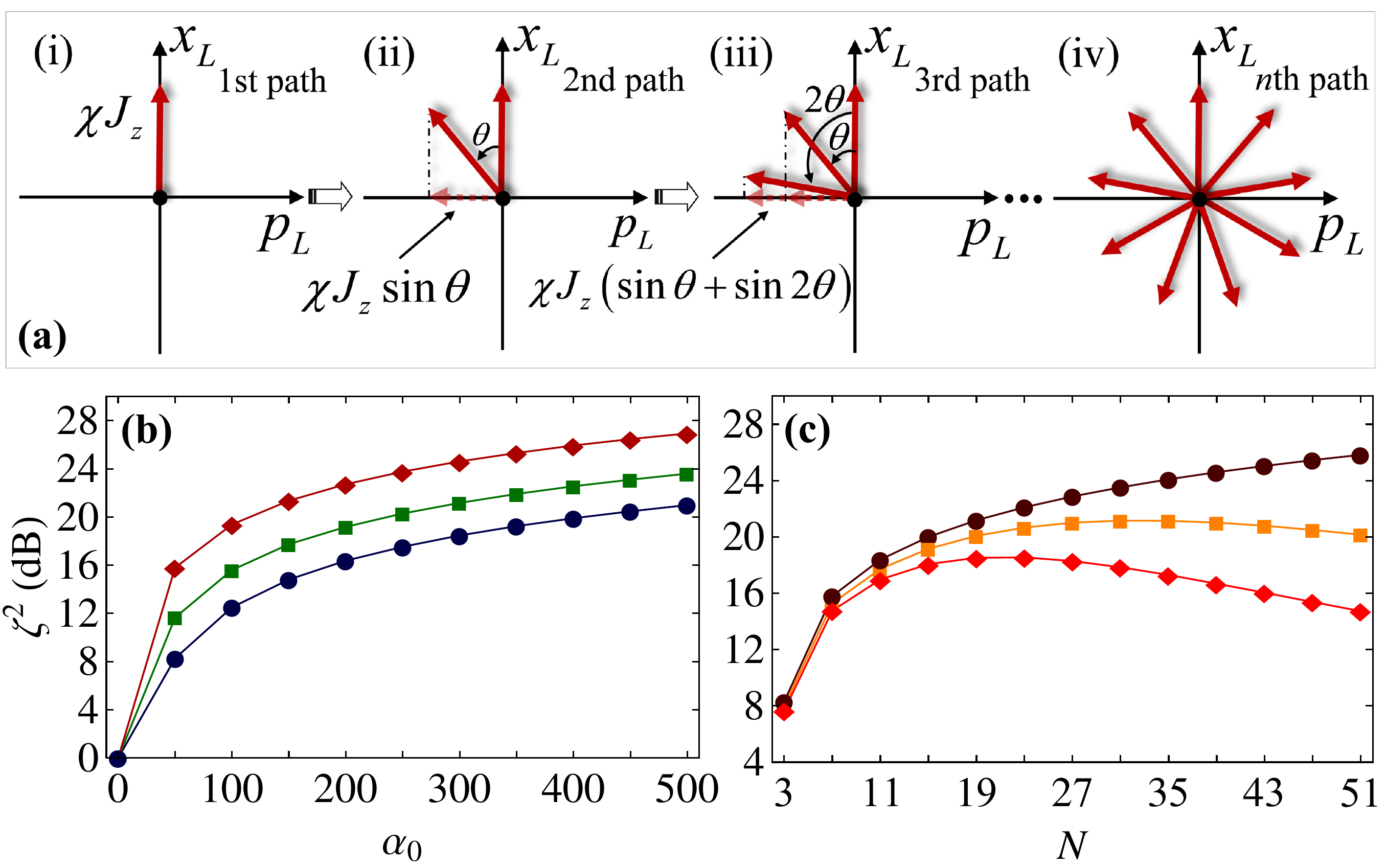}
\caption{(color online). (a) Phase-space portrait of the light's noise evolution induced by multi-pass scheme (we here depict the 9-pass as an example). During the first path, atomic noise $J_z$ is mapped onto $x_L$ ($\propto S_y$). (ii) After a polarization rotation by an angle $2\pi/N$ produced by a wave plate, the second path imprints atomic noise $J_z$ onto the $x_L$ quadrature. (iii) After another $2\pi/N$ rotation by the wave plate, $x_L$ `sees' $J_z$ again. After the ($N-1$)th rotation by the wave plate and the $N$th path, we finally arrive at (iv). The sum of the $N$ spin contribution is zero to ensure the cancellation of the spin-probe entanglement. Dashed red arrows along the $x_L$-axis are the atomic information brought back by light during each path. (b) For the case of TAT, the dependence of the squeezing $\xi^2$ on the optical depth $\alpha_0$ for different number of passes $N$, with $N=7$, $4$, $3$ from top to bottom. (c) The dependence of squeezing on the number of passes $N$, for $\alpha_0=50$, with optical loss of $0$, $0.5\%$, $1\%$ per pass from top to bottom. We assume $\phi=0$ in both (b) and (c).}
\end{figure}
  \section{Generalization to multi-pass quantum erasure}
We found that, to realize the above unitary OAT and TAT spin squeezing, the number of passes through the atomic medium does not have to be three. In fact, we can generalize the current three-pass scheme to an $N$-pass ($N\geq 3$) scheme. To illustrate how this works, we again use the phase-space portrait of the light's noise evolution as shown in Fig.4(a). For simplicity, we assume the point in the phase space for light initially is at the origin of the $x_Lp_L$ plane. Furthermore, we assume that, during each interaction, $x_L$ ($\propto S_y$) picks up only the atomic $+J_z$ contribution, which can be realized by using an optical-ring-cavity-like configuration where all the $N$ light passes go through the atomic medium along the same direction. As can be seen from Fig. 4 (a), in each path the probe light picks up information about $J_z$, which is proportional to the coupling constant $\chi$.  Similar to the three-pass scheme above, to erase the spin-probe entanglement, the sum of the spin contribution from the $N$-pass interaction should be zero [see Fig. 4(a)(iv)], which requires a rotation of the Stokes vector around the $x$-axis by an angle $\theta=2\pi/N$ after each path. From the point of view of the atoms, after the first path the atoms start to receive information about $S_z$ ($\propto p_L$) of the light in the amount $\digamma_{i}$ for the $i$th path, where $S_z$ in each pass is the projection of all the vectors (denoted by red arrows) onto the $p_L$ axis. For instance, for the second pass, the atoms obtain $S_z$ contribution $\digamma_2=\chi^2 J_z\sin\theta $ [see Fig. 4(a)(ii)], and for the third pass, $\digamma_3=\chi^2 J_z(\sin\theta+\sin2\theta) $ . Finally, for the $N$th path, we will have $\digamma_{N}=\chi^2 J_z(\sin\theta+\sin2\theta+\cdot\cdot\cdot+\sin[(N-1)\theta]) $. After all the passes, the atomic spin obtains information
\begin{eqnarray}
\Lambda=\sum\limits_{i = 2}^N {{\digamma_i}}  &=& {\chi ^2}{J_z}\sum\limits_{n = 1}^{N - 1} {(N - n)\sin \left( {n\theta } \right)} \nonumber\\
 &=& \frac{N}{2}\cot \left( {\frac{\pi }{N}} \right){\chi ^2}{J_z}.\label{Eq9-9}
\end{eqnarray}
Apparently, for $N=3$, the atomic information accumulated is $\Lambda=\sqrt{3}\chi^2 J_z/2$, which agrees with the result derived above. For large $N$, we have $\Lambda\simeq N^2\chi^2J_z/\pi\propto N^2\alpha_0$,where the enhancement factor of $N^2$ on the coupling coefficient $\chi^2$ will be reduced to $N$ after we take into account the decay of the collective spin due to spontaneously emitted light, or optical loss from the optical window or mirrors. We found that for large $N$, the squeezing factor is $\xi^2\propto1/(N\alpha_0)$. This linear enhancement with $N$ is similar to the scenario for atoms in a cavity with mirror loss, where the effective optical depth is enhanced by a factor proportional to the average number of round trips of the photon in the cavity.

We have carried out numerical calculations for different number of round trips in the presence of atomic spin loss and optical loss. In Fig.4 (b), we show in the TAT case the squeezing factor $\xi^2$ v.s. the optical depth (OD) $\alpha_0$ for three different numbers of total passes $N=3,4,7$ respectively. Here, the atomic spin loss has been considered, but the optical loss and the angle between the beam passes has been neglected, and we have optimized the laser power for each calculated data point. As can be seen, with the increase of $N$, the amount of attainable squeezing is significantly improved. In Fig.4(c), the dependence of squeezing $\xi^2$ on $N$ is shown for a relatively low optical depth $\alpha_0=50$, for different values of optical loss and with spin loss taken into account. The laser power has also been optimized for each data point. It can be seen that when the optical loss is zero (upper curve), $\xi^2$ scales linearly with $N$ for large $N$. With finite mirror loss, there is an optimal value for $N$ and it increases when the optical loss decreases. Therefore, the multi-pass approach provides a possible way to realize ultra-strong spin squeezing in low optical-depth systems, such as room temperature vapor cells. Furthermore,this multi-pass scheme should be extendable to an optical ring cavity system, where a cavity birefringence can be introduced to play the role of the waveplates here, and the amount of birefringence can be optimized according to the cavity loss to allow many successive sets of the $N$-pass quantum erasure described here, but with descending laser power.

 \section{Experimental feasibility of the three-pass scheme}
 Our TAT scheme can be implemented in both cold-atom and room-temperature atomic vapor cell systems. Here, we take the latter as an example. Consider a paraffin-coated glass cell filled with warm ${}^{87}$Rb atomic vapor, initially pumped to the ground hyperfine state $|F=2,m_F=2\rangle$ forming the CSS. The probe pulse is tuned to the D2 line ($\Gamma=3.03$ MHz) at a large blue detuning of around $\Delta=1$ GHz.
 For a cylindrical cell with diameter 2.5 cm and length of 8 cm containing $2.0\times 10^{12}$ atoms at the temperature $T=312$ K (atomic density $5.2\times 10^{10}$cm$^{-3}$), a resonant optical depth $\alpha_0=50$ can be achieved. For optimal squeezing, the decay parameter $\tilde\eta$ should be within the range of $0.10-0.30$, corresponding to a photon number per pulse on the order of $10^{14}$, with a time duration about 5 ms (spin decoherence by the wall is negligible since the typical $T_2$ time is more than 100 ms) with the power of 5 mW. Under these conditions photon loss $\epsilon$ due to atom scattering is less than $2.0\times 10^{-3}$ and thus can be neglected. For a moderate reflectivity $r_0=1\%$ \cite{PhysRevA.74.064301}, peak squeezing of $7.4$ dB is obtained at $\tilde\eta=0.26$, corresponding to the coupling constant $\kappa=2.08$. The magnetic field  added should satisfy $B=\sqrt{3}\hbar\kappa^2/(4g_F\mu_BT)$ with $g_F$ the hyperfine Land$\acute{e}$ g-factor and $\mu_B$ the Bohr magneton. For the typical duration of 5 ms, the magnetic field needed is about $0.38$ mG, corresponding to a Larmor frequency of 270 Hz. Finally, we note that, although the overlapping laser beams can form a standing wave and introduce inhomogeneous laser intensity distribution within the atomic sample, it can be averaged out by atomic motion for warm atoms, and in cold atoms can be eliminated by offsetting the frequency of the three beams to form a moving standing wave.

\section{Conclusion}
One additional benefit of our TAT scheme is the existence of the magnetic field. In most spin squeezing experiment, a magnetic field is added for spin noise detection at the Larmor frequency to avoid technical noise near DC; however, the resulted back-action diminishes squeezing and needs to be eliminated with complicated experimental configurations \cite{RevModPhys.82.1041,nphy1}. Here, since the magnetic field induced Larmor precession is built into the scheme, there is simply no back-action. The squeezing process can be understood as a continuous swapping of spin-light entanglement created by the Faraday interactions into spin-spin entanglement. Another intriguing feature of the current TAT method is that it relies on the ``simultaneous passage'' mechanism, which removes the impractical requirement of long delay lines between passages \cite{PhysRevA.74.011802} and greatly simplifies the experiment.

In summary, we have presented a novel scheme for quantum erasure which enables unitary spin squeezing. Schemes for realizing both one-axis and two-axis twisting spin squeezing of an atomic ensemble are presented. Under the Faraday-type interactions, pure SSS can be created by simply passing an optical pulse through the Larmor precessing spins for three times. Neither projective measurements nor feedback control are required.  Taking into account the noise effects imposed by several incoherent processes, our calculation shows that substantial squeezing is still obtainable under moderate experimental conditions. This proposal points to new and simple approaches to quantum erasure and unitary spin squeezing, and thus will advance technologies for quantum metrology and quantum information processing.

\begin{acknowledgments}
This work was funded by NBRPC (973 Program Grants No. 2012CB921604 and No. 2011CB921604) and NNSFC (Grant No. 11322436 and No. 11504273).
\end{acknowledgments}

\appendix
%

\section{Evolution of the spin state}
\label{Evolution of the spin state}

According to the Hamiltonian $H$ given in the main text, one may evaluate the Heisenberg equations $\partial_{t} \textbf{J} = \frac{1}{{i\hbar }}[\textbf{J},H]$ for atoms and the Maxwell-Bloch equations $(\partial_{t}+c\partial_{z})q(z,t) = \frac{1}{{i\hbar }}[q(z,t),H]$ for light \cite{PhysRevA.72.052313}, yielding
\begin{eqnarray}
\frac{d}{{dt}}{J_x}\left( t \right) &=&  - \frac{\chi }{{\sqrt T }}{J_y}\left[ {{p_L}\left( {0,t} \right) - {p_{L,\alpha }}\left( {{d_1},t} \right)} \right.\nonumber\\
&&\left. { + {p_{L,\beta }}\left( {{d_2},t} \right)} \right],\label{S1}\\
\frac{d}{{dt}}{J_y}\left( t \right) &=&  - \Omega {J_z}\left( t \right) + \frac{\chi }{{\sqrt T }}{J_x}\left[ {{p_L}\left( {0,t} \right) } \right.\nonumber\\
&&\left. {- {p_{L,\alpha }}\left( {{d_1},t} \right) + {p_{L,\beta }}\left( {{d_2},t} \right)} \right],\label{S2}\\
\frac{d}{{dt}}{J_z}\left( t \right) &=& \Omega {J_y}\left( t \right),\label{S3}\\
\left( {{\partial _t} + c{\partial _z}} \right){x_L}\left( {r,t} \right) &=& \frac{{c\chi }}{{\sqrt T }}{J_z}\left( t \right)\left[ {\delta \left( r \right) - \delta \left( {r - {d_1}} \right)\cos \alpha } \right.\nonumber\\
&&\left. { + \delta \left( {r - {d_2}} \right)\cos \beta } \right],\label{S4}\\
\left( {{\partial _t} + c{\partial _z}} \right){p_L}\left( {r,t} \right) &=&  - \frac{{c\chi }}{{\sqrt T }}{J_z}\left( t \right)\left[ {\delta \left( {r - {d_1}} \right)\sin \alpha } \right.\nonumber\\
&&\left. { - \delta \left( {r - {d_2}} \right)\sin \beta } \right].\label{S5}
\end{eqnarray}
 The first terms in (\ref{S2}) and (\ref{S3}) describe the Larmor precession of the atomic system around the $x$ axis, while the first term in (\ref{S1}) and the second term in (\ref{S2}) describe the rotation of the collective spin around the $z$ axis caused by the three successive light pass through the atomic ensemble. To simplify this set of coupled equations, we make the experimentally reasonable assumption that the spin-light coupling is weak, so that the light-induced rotation of the collective spin is small. Under this assumption, one may neglect the deviation of the macroscopic component $J_x$ from the $x$-direction after interaction [Eq. (\ref{S1}) can therefore be omitted] and
treat the operator $J_x$  by a $c$-number, $J_x\approx\langle J_x\rangle$. Next, we introduce a new position variable $\tilde r=ct-r$ which represents a
coordinate system fixed on the light pulse and allows us to
denote particular pieces on the propagating pulse easily \cite{PhysRevA.72.052313,RevModPhys.82.1041}. With this definition, the above equations become
\begin{eqnarray}
\frac{d}{{dt}}{J_y}\left( t \right) &=&  - \Omega {J_z}\left( t \right) + \frac{\chi }{{\sqrt T }}\left\langle {{J_x}} \right\rangle \left[ {{{\bar p}_L}\left( {ct,t} \right)} \right.\nonumber\\
&&\left. { - {{\bar p}_{L,\alpha }}\left( {ct - {d_1},t} \right) + {{\bar p}_{L,\beta }}\left( {ct - {d_2},t} \right)} \right],\label{S6}\\
\frac{d}{{dt}}{J_z}\left( t \right) &=& \Omega {J_y}\left( t \right),\label{S7}\\
{\partial _t}{{\bar x}_L}\left( {\tilde r,t} \right) &=& \frac{{c\chi }}{{\sqrt T }}{J_z}\left( t \right)\left[ {\delta \left( {ct - \tilde r} \right) - \delta \left( {ct - \tilde r - {d_1}} \right)\cos \alpha } \right.\nonumber\\
&&\left. { + \delta \left( {ct - \tilde r - {d_2}} \right)\cos \beta } \right],\label{S8}\\
{\partial _t}{{\bar p}_L}\left( {\tilde r,t} \right) &=&  - \frac{{c\chi }}{{\sqrt T }}{J_z}\left( t \right)\left[ {\delta \left( {ct - \tilde r - {d_1}} \right)\sin \alpha } \right.\nonumber\\
&&\left. { - \delta \left( {ct - \tilde r - {d_2}} \right)\sin \beta } \right]\label{S9},
\end{eqnarray}
where we have defined the new light quadratures $
\bar q_L \left( {\tilde r ,t} \right) = \bar q_L \left( {ct - \tilde r ,t} \right)$. Apparently, to solve the equations for atoms, it requires to first solve the equations for light. By integrating Eqs. (\ref{S8}) and (\ref{S9}) over $t$ on both sides, the Dirac delta functions are then turned into the Heaviside step functions $\Theta(\cdot)$:
\begin{eqnarray}
{{\bar x}_L}\left( {\tilde r,t} \right) &=& {{\bar x}_L}\left( {\tilde r,0} \right) + \frac{\chi }{{\sqrt T }}\left[ {{J_z}\left( {\tilde r/c} \right)\Theta \left( {t - \tilde r/c} \right)} \right.\nonumber\\
 &&- {J_z}\left( {\tilde r/c + {d_1}/c} \right)\Theta \left( {t - \tilde r/c - {d_1}/c} \right)\cos \alpha \nonumber\\
&&\left. { + {J_z}\left( {\tilde r/c + {d_2}/c} \right)\Theta \left( {t - \tilde r/c - {d_2}/c} \right)\cos \beta } \right],\nonumber\\
{{\bar p}_L}\left( {\tilde r,t} \right) &=& {{\bar p}_L}\left( {\tilde r,0} \right) - \frac{\chi }{{\sqrt T }}\left[ {{J_z}\left( {\tilde r/c + {d_1}/c} \right)} \right.\nonumber\\
 &&\times \Theta \left( {t - \tilde r/c - {d_1}/c} \right)\sin \alpha  - {J_z}\left( {\tilde r/c + {d_2}/c} \right)\nonumber\\
&&\left. { \times \Theta \left( {t - \tilde r/c - {d_2}/c} \right)\sin \beta } \right].\nonumber
\end{eqnarray}
The light quadratures $\bar p_L \left( {ct,t} \right)$, $\bar p_{L,\alpha} \left( {ct - d_{1} ,t} \right)$, $\bar p_{L,\beta} \left( {ct - d_{2} ,t} \right)$ in Eq. (\ref{S6}) can then be respectively calculated to give:
  \begin{subequations}
\label{eq:whole}
\begin{eqnarray}
{{\bar p}_L}\left( {ct,t} \right) &=& {{\bar p}_L}\left( {ct,0} \right),\label{S10a}\\
{{\bar p}_{L,\alpha }}\left( {ct - {d_1},t} \right) &=& {{\bar p}_{L,\alpha }}\left( {ct - {d_1},0} \right)\nonumber\\
 &&- \frac{\chi }{{\sqrt T }}{J_z}\left( {t - {d_1}/c} \right)\sin \alpha ,\label{S10b}\\
{{\bar p}_{L,\beta }}\left( {ct - {d_2},t} \right) &=& {{\bar p}_{L,\beta }}\left( {ct - {d_2},0} \right)\nonumber\\
 &&- \frac{\chi }{{\sqrt T }}\left[ {{J_z}\left( {t - {d_2}/c} \right)\sin \beta } \right.\nonumber\\
&&\left. { + {J_z}\left( {t - {d_2}/c + {d_1}/c} \right)\sin \left( {\alpha  - \beta } \right)} \right].\nonumber\\\label{S10c}
\end{eqnarray}
 \end{subequations}
Obviously, these three momentum quadratures are simultaneously `seen' by the spin component $J_z$. At time $t=t$, the piece $\tilde r=ct$ of the pulse enters the atomic sample for the first time, therefore it contains, as can be seen in Eq. (\ref{S10a}), no information about atoms. The situation for the piece $\xi=ct-d_1$, however, is different. Since it sits in front of $\xi=ct$ with a distance $d_1$, information of $J_z$ were marked on $\bar p_{L,\alpha} ( {ct-d_1})$ at an earlier time $c/d_1$ [see Eq. (\ref{S10b})]. The piece $\tilde r=ct-d_2$ caries information about $J_z$ at different time [the two terms in the square brackets in Eq. (\ref{S10c})], which are picked up by this piece of light during its first and second interactions with atoms. Substituting (\ref{eq:whole}) into (\ref{S6}), one obtains
\begin{eqnarray}
\frac{d}{{dt}}{J_y}\left( t \right) &=&  - \Omega {J_z}\left( t \right) + \frac{{{\chi ^2}}}{T}\left\langle {{J_x}} \right\rangle \left[ {{J_z}\left( {t - {d_1}/c} \right)\sin \alpha } \right.\nonumber\\
 &&- {J_z}\left( {t - {d_2}/c} \right)\sin \beta \nonumber\\
&&\left. { - {J_z}\left( {t - {d_2}/c + {d_1}/c} \right)\sin \left( {\alpha  - \beta } \right)} \right]\nonumber\\
 &&+ \frac{\chi }{{\sqrt T }}\left\langle {{J_x}} \right\rangle \left[ {{{\bar p}_L}\left( {ct,0} \right)} \right. - {{\bar p}_{L,\alpha }}\left( {ct - {d_1},0} \right)\nonumber\\
&&\left. { + {{\bar p}_{L,\beta }}\left( {ct - {d_2},0} \right)} \right].\label{S11}
\end{eqnarray}
Note that, in reality, the distances $d_{1,2}$ along the optical path are usually on the order of meter, which is much shorter than the spatial extension of the spatially localized light modes, that is, $c/b$ (where $b$ is assumed to be of the order of MHz). We therefore can neglect the spatial arguments of the light operators in (\ref{S11}) and finally arrive at Eq. (7) in the main text.

\section{Imperfection of the scheme}
\label{Imperfection of the scheme}

So far, we have neglected the imperfections of the setup as well as noise effects. In deriving Eq. (2), we have used the condition $\phi\rightarrow 0$, while in reality the angle $\phi$ is a small but nonzero parameter, which transforms the total Hamiltonian into:
\begin{eqnarray}
H &=& \hbar \Omega {J_x} + \frac{{\hbar \chi }}{{\sqrt T }}\left[ {{J_z}{p_L}(0) + {J_{\pi  + \phi }}{p_{L,\alpha }}({d_1})} \right.\nonumber\\
&&\left. { + {J_{2\pi  - \phi }}{p_{L,\beta }}({d_2})} \right],\label{S12}
\end{eqnarray}
 For noise effect, as mentioned in the main text, on the one hand atoms undergo dissipation due to collisional relaxation and weak excitation by the probe light, which causes decoherence of the transverse components of the spin state at a rate of $\tilde\eta/T$ . On the other hand, light also suffers losses due to reflections by
the cell walls and the incoherent scattering by atoms. Such losses, affecting both the quantum variables and
classical field, can be characterized by the reflection coefficient $\zeta$.

  By taking into account of the effect of the small angle $\phi$ and the atomic damping, the atomic evolution equations of (\ref{S6}) and (\ref{S7}) are changed into:
\begin{eqnarray}
\frac{d}{{dt}}{J_y}\left( t \right) &=&  - \Omega {J_z}\left( t \right) - \frac{{\tilde \eta }}{{2T}}{J_y}\left( t \right) + \frac{\chi }{{\sqrt T }}\left\langle {{J_x}} \right\rangle \left[ {{{\bar p}_L}\left( {ct,t} \right)} \right.\nonumber\\
 &&- \sqrt {1 - \zeta } {{\bar p}_{L,\alpha }}\left( {ct - {d_1},t} \right)\cos \phi \nonumber\\
&&\left. { + \sqrt {1 - 2\zeta } {{\bar p}_{L,\beta }}\left( {ct - {d_2},t} \right)\cos \phi } \right] + \sqrt {\frac{{\tilde \eta }}{T}} {f_{{J_y}}},\nonumber\\\label{S13}\\
\frac{d}{{dt}}{J_z}\left( t \right) &=& \Omega {J_y}\left( t \right) - \frac{{\tilde \eta }}{{2T}}{J_z}\left( t \right)\nonumber\\
 &&- \frac{\chi }{{\sqrt T }}\left\langle {{J_x}} \right\rangle \left[ {\sqrt {1 - \zeta } {{\bar p}_{L,\alpha }}\left( {ct - {d_1},t} \right)} \right.\nonumber\\
&&\left. { + \sqrt {1 - 2\zeta } {{\bar p}_{L,\beta }}\left( {ct - {d_2},t} \right)} \right]\sin \phi  + \sqrt {\frac{{\tilde \eta }}{T}} {f_{{J_z}}}.\nonumber\\\label{S14}
\end{eqnarray}
 Each time the pulse transits through the ensemble, photon scattering losses occurs. Before the next transit of the pulse through the atoms, it requires to cross two cell walls, which introduces photon reflection losses. These losses lead to the decay of the total photon number $N_{ph}$, and thus reduce the coupling strength in terms of $\chi\rightarrow\sqrt{1-(n-1)\zeta}\chi$, where $n$ accounts for $n$th interaction of light with atoms. Here, we have neglected the light reflection by the first wall. Since the input light state involved here is a coherent state, the losses due to the first crossing can always be compensated by using a more intensive pulse \cite{PhysRevA.73.062329}. The losses process also affects the quantum variables and transform (according to the main text) the light quadratures of (\ref{eq:whole}) into:
   \begin{subequations}
\label{eq:whole2}
\begin{eqnarray}
{{\bar p}_L}\left( {ct,t} \right) &=& {{\bar p}_L}\left( {ct,0} \right),\\
{{\bar p}_{L,\alpha }}\left( {ct - {d_1},t} \right) &=& \sqrt {1 - \zeta } \left[ {{{\bar p}_{L,\alpha }}\left( {ct - {d_1},0} \right)} \right.\nonumber\\
&&\left. { - \frac{\chi }{{\sqrt T }}{J_z}\left( {t - {d_1}/c} \right)\sin \alpha } \right] \nonumber\\&&+ \sqrt \zeta  {F_{L\alpha ,3}}\left( t \right),\\
{{\bar p}_{L,\beta }}\left( {ct - {d_2},t} \right) &=& \sqrt {1 - 2\zeta } \left\{ {{{\bar p}_{L,\beta }}\left( {ct - {d_2},0} \right)} \right.\nonumber\\
 &&- \frac{\chi }{{\sqrt T }}\left[ {{J_z}\left( {t - {d_2}/c} \right)\sin \beta } \right.\nonumber\\
&&\left. {\left. { - {J_{\pi  + \phi }}\left( {t - {d_2}/c + {d_1}/c} \right)\sin \left( {\alpha  - \beta } \right)} \right]} \right\}\nonumber\\
 &&+ \sqrt {2\zeta } {F_{L\beta ,6}}\left( t \right),
\end{eqnarray}
 \end{subequations}
where we have defined the collective light noise operators $
{F_{L\vartheta ,m}} = {f_{Lp,m}}\cos \vartheta  - {f_{Lx,m}}\sin \vartheta$.
By inserting Eqs. (\ref{eq:whole2}) into (\ref{S13}) and (\ref{S14}), one may obtain the new
evolutions for atoms:
\begin{eqnarray}
{\partial _t}\left( {\begin{array}{*{20}{l}}
{{J_y}\left( t \right)}\\
{{J_z}\left( t \right)}
\end{array}} \right) &=& {C_1}\left( {\begin{array}{*{20}{l}}
{{J_y}\left( t \right)}\\
{{J_z}\left( t \right)}
\end{array}} \right) + {C_2}\left( {\begin{array}{*{20}{l}}
{{{\bar x}_L}\left( {ct,0} \right)}\\
{{{\bar p}_L}\left( {ct,0} \right)}
\end{array}} \right)\nonumber\\
 &&+ {C_3}\left( {\begin{array}{*{20}{l}}
{{F_{L\alpha ,3}}\left( t \right)}\\
{{F_{L\beta ,6}}\left( t \right)}
\end{array}} \right) + \sqrt {\frac{\eta }{T}} \left( {\begin{array}{*{20}{l}}
{{f_{{J_y}}}(t)}\\
{{f_{{J_z}}}(t)}
\end{array}} \right).\label{eq21a}\nonumber\\
\end{eqnarray}
Here the coefficients matrix $C_1$, $C_2$, and $C_3$ can easily be calculated to give:
\begin{widetext}
\begin{eqnarray}
{C_1} &=&  - \frac{1}{T}\left( {\begin{array}{*{20}{c}}
{\frac{{\tilde \eta }}{2} - \frac{{{\kappa ^2}}}{2}\left( {1 - 2\zeta } \right)\sin 2\phi \sin \left( {\alpha  - \beta } \right)}&{\Omega T - \cos \phi {\kappa ^2}\left[ {{S_ - } - \left( {1 - 2\zeta } \right)\cos \phi \sin \left( {\alpha  - \beta } \right)} \right]}\\
{ - \Omega T + {\kappa ^2}\left( {1 - 2\zeta } \right)\sin {\phi ^2}\sin \left( {\alpha  - \beta } \right)}&{\frac{{\tilde \eta }}{2} - \sin \phi {\kappa ^2}\left[ {{S_ + } + \left( {1 - 2\zeta } \right)\cos \phi \sin \left( {\alpha  - \beta } \right)} \right]}
\end{array}} \right),\\
{C_2} &=&  - \frac{\chi }{{\sqrt T }}\left\langle {{J_x}} \right\rangle \left( {\begin{array}{*{20}{c}}
{-\cos \phi {S_ - }}&{ - 1 + \cos \phi {C_ - }}\\
{\sin \phi {S_ + }}&{\sin \phi {C_ + }}
\end{array}} \right),\\
{C_3} &=&  - \frac{\chi }{{\sqrt T }}\left\langle {{J_x}} \right\rangle \left( {\begin{array}{*{20}{c}}
{\cos \phi \sqrt {\left( {1 - \zeta } \right)\zeta } }&{ - \cos \phi \sqrt {\left( {1 - 2\zeta } \right)2\zeta } }\\
{\sin \phi \sqrt {\left( {1 - \zeta } \right)\zeta } }&{\sin \phi \sqrt {\left( {1 - 2\zeta } \right)2\zeta } }
\end{array}} \right),
\end{eqnarray}
\end{widetext}
with $S_\pm={\left( {1 - \zeta} \right)\sin\alpha  \pm \left( {1 - 2\zeta} \right)\sin\beta },C_\pm={\left( {1 - \zeta} \right)\cos\alpha  \pm \left( {1 - 2\zeta} \right)\cos\beta }$. Eq. (\ref{eq21a}) can then be directly solved to yield the input-output relation for atoms
\begin{eqnarray}
\left( {\begin{array}{*{20}{l}}
{J_y^{out}}\\
{J_z^{out}}
\end{array}} \right) &=& C\left( T \right)\left( {\begin{array}{*{20}{l}}
{J_y^{in}}\\
{J_z^{in}}
\end{array}} \right) + C\left( T \right)\int_0^T {d\tau } {C^{ - 1}}\left( \tau  \right)\nonumber\\
 &&\times \left[ {{C_2}\left( {\begin{array}{*{20}{l}}
{{{\bar x}_L}\left( {c\tau ,0} \right)}\\
{{{\bar p}_L}\left( {c\tau ,0} \right)}
\end{array}} \right) + {C_3}\left( {\begin{array}{*{20}{l}}
{{F_{L\alpha ,3}}\left( \tau  \right)}\\
{{F_{L\beta ,6}}\left( \tau  \right)}
\end{array}} \right)} \right.\nonumber\\
&&\left. { + \sqrt {\frac{\eta }{T}} \left( {\begin{array}{*{20}{l}}
{{f_{{J_y}}}\left( \tau  \right)}\\
{{f_{{J_z}}}\left( \tau  \right)}
\end{array}} \right)} \right],\label{S20}
\end{eqnarray}
where $C(t)=e^{C_1t}$. Note that,
unlike the ideal case, now the light quadratures in (\ref{S20}) can not freely be canceled, since the adjustment of the phase shifts $\alpha$ and $\beta$, on the other hand, will amplify the noise effect of light. As a result, there exists an optimal choice of $\alpha$ and $\beta$, with which the amount of squeezing can be maximized.
\bibliography{ref}
\end{document}